\documentclass[a4paper, amsfonts, amssymb, amsmath, reprint, nofootinbib, twoside, superscriptaddress]{revtex4-1} 
\usepackage[left=1.8cm, right=1.8cm, top=1.8cm, bottom=2.0cm]{geometry}
\usepackage[english]{babel}
\usepackage[utf8]{inputenc}
\usepackage[colorinlistoftodos, color=green!40, prependcaption]{todonotes}
\usepackage{amsthm}
\usepackage{mathtools}
\usepackage{physics}
\usepackage{xcolor}
\usepackage{graphicx}
\usepackage{adjustbox}
\usepackage{placeins}
\usepackage[T1]{fontenc}
\usepackage{lipsum}
\usepackage{csquotes}
\usepackage[colorlinks,allcolors=blue!50!black]{hyperref} 
\usepackage{multirow}
\usepackage{amssymb}
\usepackage{pifont}

\begin{document}

\title{Rapid Discovery of Stable Materials by Coordinate-free Coarse Graining}

\author{Rhys E. A. Goodall}
    \affiliation{Department of Physics, University of Cambridge, United Kingdom}

\author{Abhijith S. Parackal}
    \affiliation{Department of Physics, Chemistry and Biology, Linköping University, Sweden}

\author{Felix A. Faber}
    \affiliation{Department of Physics, University of Cambridge, United Kingdom}

\author{Rickard Armiento}
    \email[Correspondence email address: ]{rickard.armiento@liu.se}
    \affiliation{Department of Physics, Chemistry and Biology, Linköping University, Sweden}

\author{Alpha A. Lee}
    \email[Correspondence email address: ]{aal44@cam.ac.uk}
    \affiliation{Department of Physics, University of Cambridge, United Kingdom}

\begin{abstract}

A fundamental challenge in materials science pertains to elucidating the relationship between stoichiometry, stability, structure, and property. Recent advances have shown that machine learning can be used to learn such relationships, allowing the stability and functional properties of materials to be accurately predicted. However, most of these approaches use atomic coordinates as input and are thus bottlenecked by crystal structure identification when investigating novel materials. Our approach solves this bottleneck by coarse-graining the infinite search space of atomic coordinates into a combinatorially enumerable search space. The key idea is to use Wyckoff representations -- coordinate-free sets of symmetry-related positions in a crystal -- as the input to a machine learning model. Our model demonstrates exceptionally high precision in discovering new theoretically stable materials, identifying \textcolor{black}{1,569} materials that lie below the known convex hull of previously calculated materials from just 5,675 \textit{ab-initio} calculations. Our approach opens up fundamental advances in computational materials discovery.

\end{abstract}


\maketitle

\section{Introduction}
Finding a needle in a haystack is often used as an analogy for materials discovery. Only a small proportion of viable material compositions (believed to be of the order $\mathcal{O}(10^{10})$  \cite{davies2016computational}) will have thermodynamically stable polymorphs that are experimentally accessible.
Most approaches to tackle this challenge focus on predictive models for materials properties -- metaphorical sieves that filter out the hay.
Here we seek an alternative approach: Can we significantly cut down materials space by changing how we represent materials -- making most of the hay disappear?

Our approach is motivated by a concept ubiquitous in science: coarse-graining.
Taking molecular chemistry, for example, chemists typically build intuitions about chemical properties using molecular graphs.
Molecular graphs are a coarse-grained representation of molecules, with each graph corresponding to a unique ensemble of atomic coordinates.
Searching in the enumerable space of molecular graphs, as opposed to the innumerable space of possible atomic coordinates, has enabled the development of powerful computational tools \cite{duvenaud2015convolutional, vamathevan2019applications} as well as efforts that exhaustively enumerate chemical space \cite{ruddigkeit2012enumeration, reymond2015chemical}.

In materials science, however, an analogous coarse-grained representation of crystal structures is missing.
Thus, we are left confronting the innumerable search space problem.
Composition-based approaches can somewhat overcome this challenge \cite{meredig2014combinatorial, ward2016general, goodall2020predicting, wang2021compositionally} but do so at the cost of discarding all information about the crystal structures of the materials being considered.
As such, either extensive computational crystal structure searching or lab-based experiments are required to validate predictions.

One avenue to manoeuvre around this challenge has been to explore restricted classes of structure prototypes using novel descriptors, e.g. Perovskites \cite{schmidt2017predicting, liu2020screening, jain2018wyckoff}, quaternary Heuslers \cite{kim2018machine}, or Elpasolites \cite{faber2016machine}.
Specifying the prototype avoids the need for crystal structure searching, empowering more extensive screening campaigns as the computational cost of validation is greatly reduced.

In this paper, we introduce an approach that generalises these prototype-restricted models by considering Wyckoff representations -- coordinate-free sets of symmetry-related positions in a crystal.
This framework allows us to develop accurate machine learning models for materials discovery tasks where the relaxed crystal structure is \textit{a priori} unknown.

We first test the ability of our model to identify novel stable materials across a diverse range of chemistries, showing that it has a precision $\sim$3 times larger than state-of-the-art coordinate-free methods based on elemental substitutions \cite{hautier2011data, wang2021predicting}. We then evaluate the performance of our model in identifying stable structures within phase diagrams with diverse structures, showing that our model finds low energy structures in the phase diagram with $\sim$5 times lower computational effort. Finally, we develop a materials exploration pipeline that, starting from an initial nucleus of known materials, screens nearby materials space and allows the efficient discovery of new stable materials. We identify \textcolor{black}{1,569} hitherto unknown materials that are below the known convex hull of previously calculated materials from just 5,675 \textit{ab-initio} calculations.

\section{Results}

\begin{figure*}
    \centering
    \includegraphics[width=0.9\textwidth]{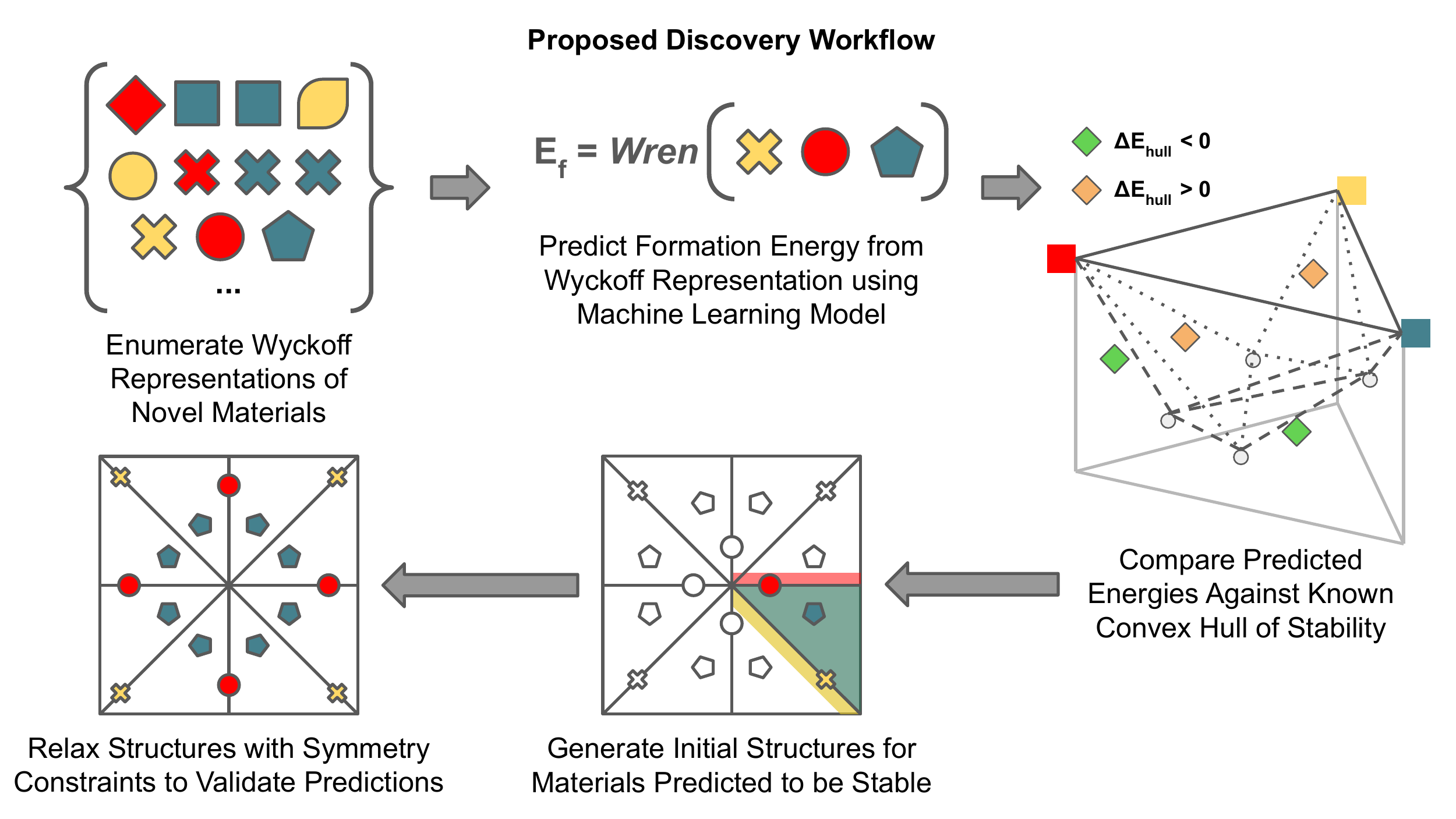}
    \caption{\label{fig:toy} \textbf{Coarse-graining materials space using Wyckoff representations enables efficient data-driven materials discovery.}
    A machine learning powered materials discovery workflow that takes advantage of the benefits of the proposed Wyckoff representation.
    The workflow uses a machine learning model to predict formation energies for candidate materials in an enumerated library of Wyckoff representations (shapes are used to denote different Wyckoff positions and colours to denote different element types).
    These predicted formation energies are then compared against the known convex hull of stability.
    Structures satisfying the required symmetries are then generated and relaxed for materials predicted to be stable.
    The calculated energies of the relaxed structures can then be compared against the known convex hull to confirm whether the candidate is stable.
    }
\end{figure*}

\subsection{Wyckoff Representation Regression}

Building an accurate machine learning model hinges on identifying model inputs that are sufficiently informative to allow the target variable to be predicted.
However, for a machine learning model to be useful in practice, these inputs need to be significantly cheaper to obtain than the cost of labelling data.
In the context of materials discovery, previous works have shown that virtual screening workflows based on Kohn-Sham density functional theory (DFT) can be used to identify novel functional materials \cite{ceder1998identification, jain2016computational}.
Separately it has been shown that accurate machine learning models can be built for the formation energies of inorganic crystals calculated via DFT using the DFT-relaxed crystal structure as the model input \cite{schutt2018schnet, xie2018crystal, chen2019graph, park2020developing}. Inference using these models is significantly cheaper than the DFT calculations they approximate, but sadly their application to materials discovery is circular because arriving at a DFT-relaxed structure necessitates calculating the energy using DFT multiple times.

Several groups have therefore proposed to use composition-based inputs \cite{meredig2014combinatorial, ward2016general, goodall2020predicting, wang2021compositionally}, which avoid the upfront need for structure identification.
However, the composition is not expressive enough to differentiate polymorphs.
This is a significant shortcoming as different polymorphs can have radically different properties, most famously the example of diamond and graphite.
As such, we turn to model inputs that can distinguish polymorphs whilst also avoiding the cost of DFT.
Such models can be used to triage which DFT calculations are carried out in materials discovery workflows, allowing for more efficient use of computational resources.

In crystallography, one way to completely specify the crystal structure of a material is via a combination of: 1) the spacegroup of the structure, 2) the dimensions of its unit cell, and 3) a set of Wyckoff positions with the elements that sit on them.
The Wyckoff positions describe sites that map onto equivalent sites under the symmetry transformations of the given spacegroup \cite{wyckoff1922analytical}.
As a consequence, a single Wyckoff position can encode the positions of multiple atoms.
To construct model inputs from sets of Wyckoff positions, we discard the information about the exact positions and lattice parameters.
In the resulting coordinate-free representation, the Wyckoff representation, each Wyckoff position is simply labelled by a Wyckoff letter and the element at that position.
Consequently, as the Wyckoff representation is discrete, it is possible to computationally enumerate Wyckoff representations that represent candidate materials for use in screening campaigns.

The procedure of obtaining the Wyckoff representation from a crystal structure can be viewed as a coarse-graining process that takes us from an unsymmetrised initial parameter space of size $4N + 6$, through the symmetrised Wyckoff position space of maximum size $5M + 6$, to the much smaller coordinate-free space of Wyckoff representations with size $2M$, where $N$ is the number of sites in the unit cell and the corresponding number of Wyckoff positions $M$ satisfies $M \leq N$.
The back mapping from the coarse-grained Wyckoff representation to the full structure can often be satisfactorily obtained via a single symmetry constrained DFT-relaxation of a prototype structure (see \autoref{fig:toy}).

To use the Wyckoff representation as the input for a machine learning model, we formulate the task of property prediction as a multi-set regression problem.
A message-passing neural network architecture based on the \textit{Roost} architecture \cite{goodall2020predicting} is used to do this -- the \textit{Roost} model performs materials property prediction via set regression on the weighted set of elements in a material's composition.

The principal idea behind the model architecture is to embed the coordinate-free Wyckoff positions of a given material into a vector space.
The representations in this embedded space are then updated via message passing operations that consider all directed pairwise combinations of members in the multi-set.
The messages propagate contextual information between Wyckoff positions leading to the emergence of material-specific representations.
These message passing stages are repeated multiple times before a permutation invariant pooling operation is applied to the multi-set to get a fixed-length representation.
As the labelling of Wyckoff positions includes several choices of setting, we carry out on-the-fly augmentation of equivalent Wyckoff representations. We then average the fixed-length representations for these equivalent inputs to ensure invariance to this choice.
These averaged fixed-length representations are then fed into a feed-forward output neural network that returns the models predictions.

This work focuses primarily on models that predict the formation energy of inorganic crystalline materials, although the proposed framework and inputs are applicable to any material property.
We call the proposed model \textit{Wren} (\textbf{W}yckoff \textbf{Re}presentation regressio\textbf{N}).
Throughout this work, we train Deep Ensembles consisting of 10 \textit{Wren} models starting from different random initialisations \cite{lakshminarayanan2017simple}, allowing us to estimate the model's uncertainty as well as providing better point estimates.
Details of the \textit{Wren} architecture and the hyper-parameters used are given in the Supplementary Information.

\subsection{Selecting Stable Materials from Diverse Chemical Space}

To accelerate the screening of materials space for novel stable materials, a model must reduce the expected number of calculations needed to find a candidate below the known convex hull (here taken to be the convex hull of the MP data set before cleaning). We first assess the ability of the model to generalise across materials space to unseen combinations of elements.

To do this, we consider two data sets: (i) The Materials Project (MP) database \cite{jain2013commentary} which is a highly curated database of high-throughput DFT calculations.
At the time of access, the Materials Project database contains approximately 140k crystal structures.
We apply a canonicalisation and cleaning treatment \textcolor{black}{(see Materials and Methods)} that leaves a final MP data set containing approximately 105k distinct materials; (ii) The WBM dataset, obtained from \cite{wang2021predicting}. The dataset contains calculated energies and properties of a large number of crystal structures that were generated through the substitution of elements in known crystal structures from MP with chemically similar elements \cite{glawe2016optimal}. As such, the WBM dataset chemically extrapolates from the MP dataset. After de-duplication and cleaning the WBM data set contains approximately \textcolor{black}{215k} materials.

We make predictions for the formation energies of the materials contained in the WBM data set using a \textit{Wren} model trained on the MP data set.
We then assess how well the \textit{Wren} model selects potentially stable materials from the WBM data set.
The relevant metrics are; the prevalence -- the proportion of materials below the known convex hull (actual positives), the precision -- how many of the predictions of potentially stable materials are correct (i.e. the ratio of true predicted positives to the total predicted positives), and the recall -- how many of the actual materials below the known convex hull are found (i.e. the ratio of true predicted positives to actual positives).
In this setup, the ratio of the precision and the prevalence gives the enrichment factor or degree of acceleration.
Enrichment factors are frequently reported for virtual screening campaigns in drug discovery applications \cite{bender2005discussion, huang2006benchmarking}.

The precision using the \textit{Wren} model to triage calculations is \textcolor{black}{38\%}.
Consequently, given that the prevalence of theoretically stable materials in the WBM data set is \textcolor{black}{15\%}, using \textit{Wren} leads to an enrichment factor of 2.5.
As enrichment here is computed with respect to the active search strategy of Ref. \cite{wang2021predicting} this translates into a significant improvement in efficiency over random or exhaustive search strategies as our improvements compound multiplicatively with theirs.
Consequently, triaging screening workflows based on \textit{Wren} should enable more materials below the known hull to be identified with limited computational resources.
We also observe a high recall of \textcolor{black}{76\%}, meaning that \textit{Wren} misses relatively few potentially stable materials.

\begin{figure}[!t]
    \centering
    \includegraphics[width=0.49\textwidth]{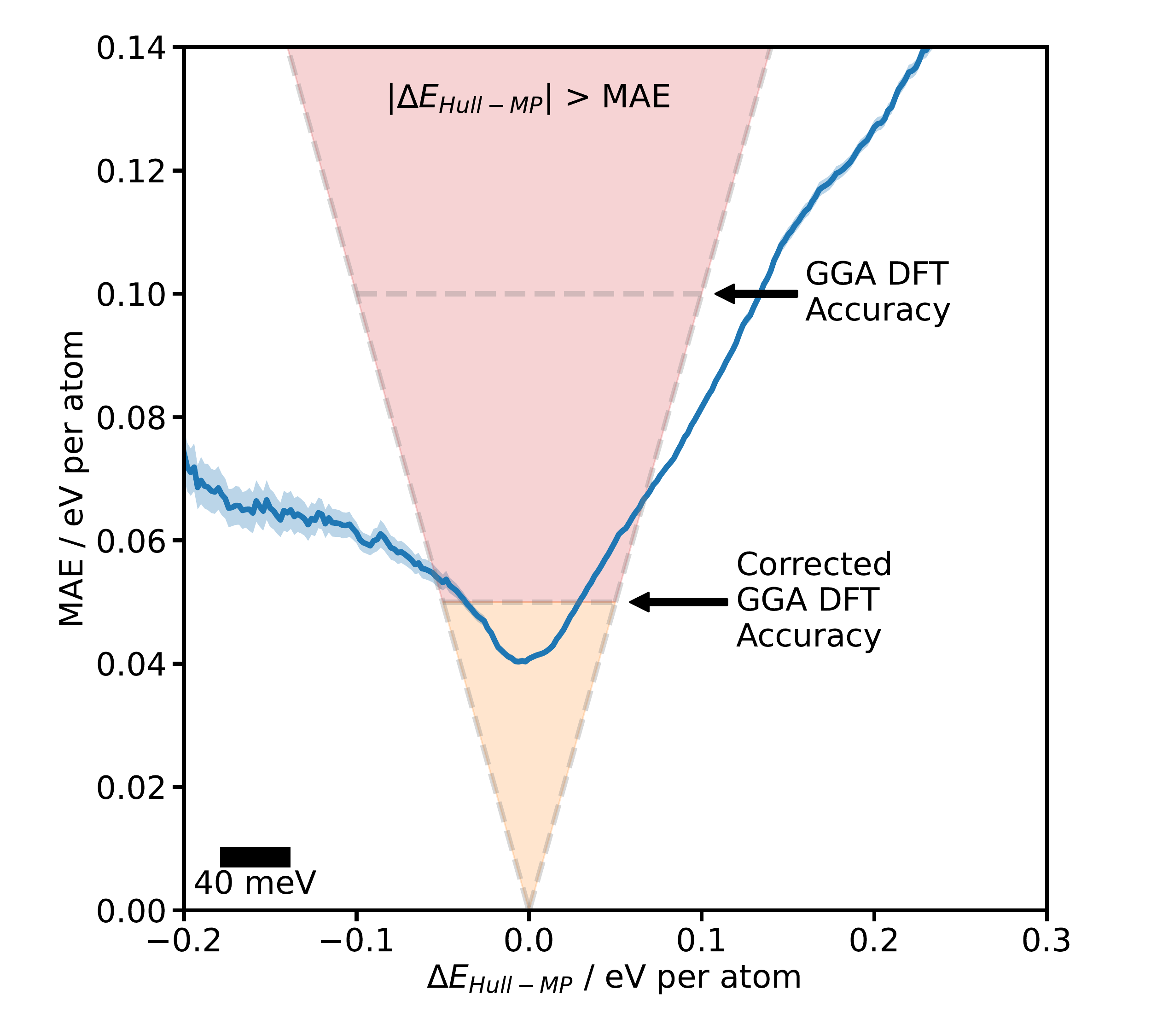}
    \caption{
    \textbf{\textit{Wren}'s average error is below DFT error in the region around the stability threshold.}
    Rolling mean absolute error on the WBM data set as the energy to the convex hull is varied for \textit{Wren} model.
    A scale bar is shown for the windowing period of 40 meV per atom used when calculating the rolling average.
    The standard error in the mean is shaded around each curve.
    The highlighted V-shaped region shows the area in which the average absolute error is greater than the energy to the known convex hull -- this is the region where the model is most at risk of miss-classifying structures.
    In the majority of this region\textit{Wren}'s accuracy is well below the 100 meV per atom threshold considered to be the accuracy of semi-local DFT across diverse chemistries \cite{kirklin2015open} and comparable to the $\sim50$ meV per atom threshold characteristic of fitted correction schemes \cite{stevanovic2012correcting, friedrich2019coordination, wang2021corrections}.
    }
    \label{fig:rolling}
\end{figure}

\begin{figure*}[!th]
    \centering
    \includegraphics[width=0.98\textwidth]{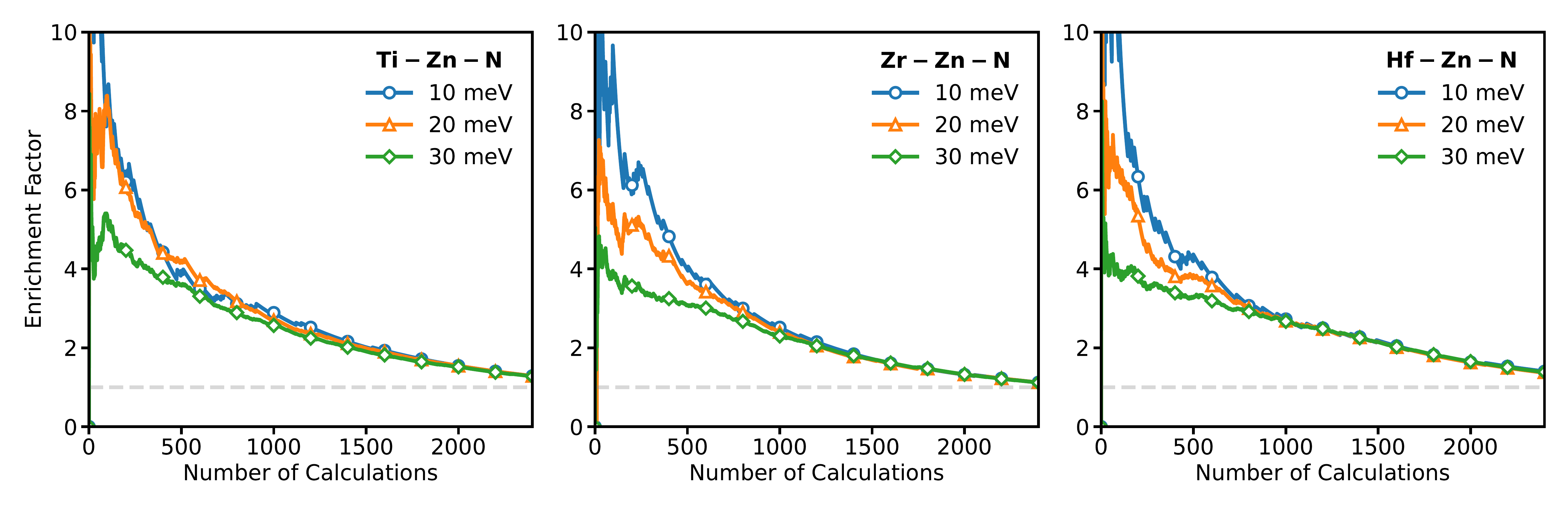}
    \caption{
    \textbf{\textit{Wren} accelerates the recovery of low energy structures in unseen chemical systems.}
    The figures show how the enrichment factor varies as we use \textit{Wren} to direct exploration of the Ti-Zn-N, Zr-Zn-N and Hf-Zn-N chemical systems.
    The enrichment factor is the ratio of candidates found satisfying a given triage criterion to the number we would expect to find via a random search.
    The enrichment factor is plotted for candidates within 10, 20 and 30 meV per atom from the convex hull of the full explored system.
    A light-grey guideline is included to show the performance expected from a random model -- an enrichment factor of 1.
    The plots demonstrate that using \textit{Wren} leads to a significant degree of early enrichment of low energy structures.
    }
    \label{fig:zero-taata}
\end{figure*}

The screening performance of the model can be tuned by adjusting our triage criteria.
For example, an alternative triage criterion would be to require that $\Delta \hat{E}_\text{Hull-Pred} + \hat{\sigma} < 0$ where $\Delta \hat{E}_\text{Hull-Pred}$ is the predicted distance of a candidate material from the known convex hull and $\hat{\sigma}$ is the predictive uncertainty of the model.
This uncertainty adjusted criterion encourages the model to suggest candidates it is more certain about, leading to an increased precision of 53\%. The enrichment factor for the uncertainty adjusted criterion is \textcolor{black}{3.5}.
Consequently, the choice of triage criteria should depend on the aims of a given workflow -- the expected opportunity cost of false negatives vs false positives, the availability of experimental or computational resources, and how easy it is to expand the candidate pool.


The strong performance of \textit{Wren} can be explained by looking at how the mean absolute error changes as a function of the distance from the known convex hull.
\autoref{fig:rolling} shows that near to the stability threshold, $\Delta E_\text{Hull-MP} = 0$, \textit{Wren} makes highly accurate predictions of the formation energy.

More significant errors are seen for materials far above and far below the hull.
However, in these regions, the average error is less than the energy to the convex hull, meaning that the model's classifications are still reliable.
The large errors far above the hull are due to the routine underestimation of the formation energy of unstable structures.
This underestimation is a manifestation of a bias in the MP data set towards structures with low formation energies.
The bias arises from the fact that large numbers of the initial structures in the MP data set are sourced from the ICSD \cite{bergerhoff1987crystallographic}.
This result highlights the importance of negative examples for building generally applicable machine learning models \cite{raccuglia2016machine, schmidt2019recent, horton2021promises}.

\subsection{Exploration of Unseen Tertiary Phase Diagrams}

From an applications perspective, researchers are often interested in exploring a single or small number of chemical systems that have not previously been studied.
Typical approaches for computationally mapping out the convex hull of novel chemical systems \cite{pickard2011ab, wang2012calypso} are highly expensive, often requiring thousands of structures to be relaxed.

To evaluate the ability of \textit{Wren} \textcolor{black}{to assist when} mapping the phase diagrams of \textcolor{black}{targeted chemical systems}, we consider the TAATA data set \cite{tholander2016strong}, consisting of 3 highly sampled ab-initio phase diagrams for the Hf-Zn-N, Ti-Zn-N and Zr-Zn-N ternary systems. The ternary systems studied in the TAATA data set were investigated for their potential in piezoelectric devices and energy harvesting applications. \textcolor{black}{We focus on ternary systems due to the fact that whilst crystal structure prediction approaches such as \cite{pickard2011ab, wang2012calypso} work very well for unary and binary systems, there has been less work applying these methods to ternary systems \cite{kvashnin2020computational, lu2021ab, di2022first} due to the combinatorial explosion in the number of candidates that need to be relaxed per phase diagram to obtain reliable results -- often in excess of 10,000 relaxations are carried out for each chemical system.}

The TAATA data set contains a diverse range of stable and unstable structures for each composition \textcolor{black}{(Full details about the construction of the TAATA data set are given in the Supplementary Information as well as in \cite{tholander2016strong}).
\textcolor{black}{After applying a canonicalisation and cleaning treatment (See Materials and Methods)}, we are left with 3,104 entries over 523 compositions in the Ti-Zn-N phase diagram, 2,711 entries over 453 compositions in the Zr-Zn-N phase diagram, and 3,381 entries over 596 compositions in the Hf-Zn-N phase diagram.}

We trained \textit{Wren} on the MP data set but excluded all the tertiary compounds from the Ti-Zn-N, Zr-Zn-N and Hf-Zn-N chemical systems. This model was then used to predict the energies of tertiary compounds in the TAATA data set. \textcolor{black}{These formation energy predictions were then used to construct hypothetical convex hulls for the Ti-Zn-N, Zr-Zn-N and Hf-Zn-N chemical systems (See Supplementary Information).} Figure \ref{fig:zero-taata} shows how \textcolor{black}{selecting which relaxations to carry out based on the predicted distances to the hypothetical convex hulls constructed using the \textit{Wren} model's predictions can accelerate phase diagram exploration. To quantify this effect we} look at the enrichment factor as a function of the number of calculations. The enrichment factor describes the ratio between the number of candidates found satisfying a target criterion when employing a given virtual screening strategy and the number of positive candidates that hypothetically would have been found if the candidates were screened randomly.
Considering materials within 20 meV per atom of the \textcolor{black}{DFT-calculated} convex hull as our target criteria, we see that after 250 calculations we have a high enrichment factor of 5.4 in the Ti-Zn-N chemical system, 5.1 for the Zr-Zn-N chemical system, and 4.5 for the Hf-Zn-N chemical system when using the \textit{Wren} model, i.e. we are saving 4.5-5.4x compute resource compared to a random search.

\begin{figure}
    \centering
    \includegraphics[width=0.49\textwidth]{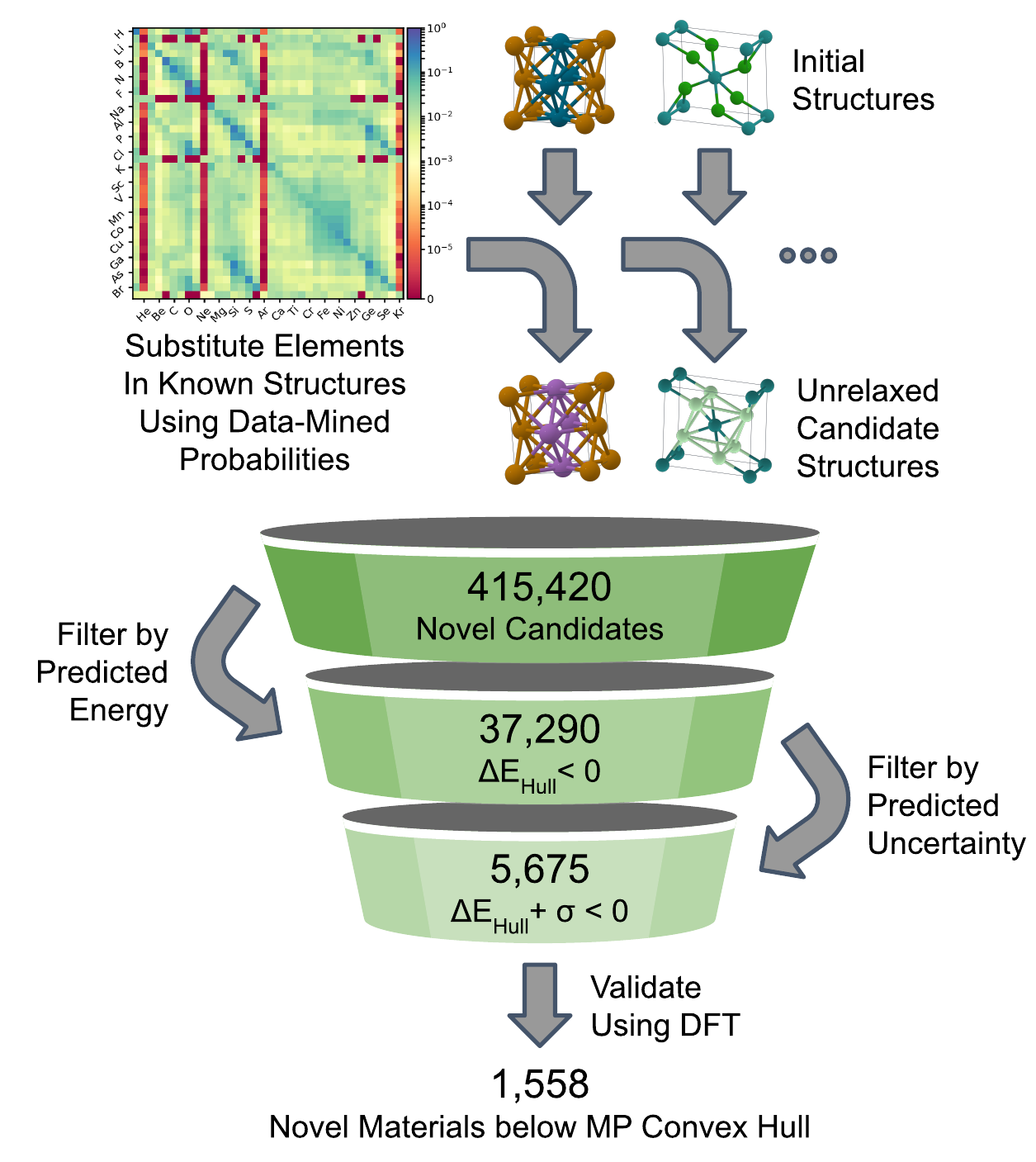}
    \caption{
        \label{fig:funnel}
        \textbf{\textit{Wren} enables automated computational prospecting of new stable materials.}
        Data-mined substitution probabilities are used to generate novel candidates for screening.
        A heat-map of the data-mined log substitution probabilities for the first 36 main group elements is shown in the top left.
        The matrix captures known chemical trends, for example, that halogens can often be substituted for each other in crystal structures.
        Using the Wyckoff representation regression allows far more unrelaxed candidates to be considered than possible in conventional DFT-led high-throughput workflows.
        The funnel diagram shows the number of unrelaxed candidates that pass the different stability criteria \textcolor{black}{when filtering based on the predictions of the \textit{Wren} model}.
        In total 4,721 out of 5,675 validation calculations completed.
        Of these \textcolor{black}{1,569} were below the known convex hull, giving a precision of 33\% amongst the completed calculations.
    }
\end{figure}

\subsection{Computational Prospecting for Novel Stable Materials}
\label{sec:pros}

Having established the promise of Wyckoff representation regression in predicting the stability of unseen materials, we deploy \textit{Wren} on the prospective challenge of discovering new theoretically stable materials.
For this stage, we trained \textit{Wren} on the union of the MP and WBM data sets.
This combined data set contains approximately 322k materials after canonicalisation and cleaning.
We randomly sampled 5\% of the data set to use as a test set and trained on the remaining 95\%.
The resulting model has a mean absolute error of 31 meV per atom on this test set, which is below the commonly quoted chemical accuracy level of 1 kcal per mol (43 meV per atom) \cite{faber2017prediction}.
The model's accuracy as a function of training set size follows a power-law relationship (see Supplementary Information).
Reassuringly, the model does not appear to saturate in performance, suggesting that the representation is rich enough and further increases in model performance can be unlocked given more data \cite{muller1996numerical, faber2015crystal, faber2018alchemical}.

Whilst the coarse-grained space of Wyckoff representations is computably enumerable and far smaller than the infinite space of atomic coordinates, attempting materials discovery by exhaustively screening all possible Wyckoff representations \textcolor{black}{for general chemistries} is computationally inefficient as the prevalence of stable materials remains vanishingly low even in the coarse-grained space. \textcolor{black}{To effectively tackle the task of materials discovery in general chemistries, it is necessary to first construct a design space with a higher expected prevalence \cite{kim2018machine}. To do this }we draw inspiration from previous work \cite{hautier2011data, wang2021predicting} and generate candidates for screening by making elemental substitutions in crystal structures that are near to the known convex hull. \textcolor{black}{Focusing on the use machine learning to accelerate workflows tackling general chemistries is particularly compelling as false positive and false negative data produced in such workflows can subsequently be used to retrain the model. Inclusion of a diverse range of negative examples in the training data is key to improving performance in targeted exhaustive workflows, such as in Section C, where design space enrichment is not applicable.}

To obtain our substitution probabilities, we extracted 39,164 ordered structures from the ICSD \cite{bergerhoff1987crystallographic, hellenbrandt2004inorganic} and binned them according to their Wyckoff representations.
Within each prototype, all pairs of structures are compared and we count which element substitutions (including self substitutions) would be needed to change one structure into the other \cite{glawe2016optimal}.
We only consider substitutions where all Wyckoff positions sharing one element-type are changed simultaneously and not per position substitutions.
Once normalised, the rows of the count matrix can be interpreted as substitution probabilities for each element.

Using these data-mined probabilities, we generated a screening library of materials by substituting different elements into structures taken from the MP data set.
We only consider initial structures from the MP data set with energies above the convex hull less than 100 meV per atom.
This choice of this threshold means that we should be including most metastable structures within the MP data set.
We consider 10 different substitutions for each initial structure.
Candidates that have the same composition as materials already present in the union of the MP and WBM data sets are removed from the library.
Lanthanide and actinide-based materials and materials containing Noble elements were also excluded.
This workflow produced a screening library of approximately 415k candidates.
\textcolor{black}{The size of the screening library can be readily increased by considering more elemental substitutions per structure.}

Despite constraining our screening set to be close to known materials, it is likely that we are still asking the model to make predictions in areas of materials space where it lacks support from the training data.
As shown on the WBM data set, uncertainty estimation allows us to reduce the risk in our materials screening process by factoring our model's uncertainty into our triage criterion.
For simplicity we use the same simple uncertainty adjusted criterion considered previously; $\Delta \hat{E}_\text{Hull-Pred} + \hat{\sigma} < 0$.
In total 5,675 candidates satisfied this screening criterion (See \autoref{fig:funnel}).

Validation with DFT resulted in 4,721 completed calculations \textcolor{black}{across 4,464 unique compositions. Of these, 1,569 structures were confirmed to be below the convex hull of the MP data set, 1,369 below the convex hull of the union of the MP and WBM data sets.} Therefore, the precision for the completed calculations was 33\% \textcolor{black}{with respect to the MP convex hull}, confirming the workflow's ability to accelerate materials discovery.
\textcolor{black}{Although direct comparisons are not strictly permissible, as previous prospective searches using machine learning have been restricted to single prototypes, the Wyckoff representation based approach presented significantly surpasses previously reported precision's of 4\% \cite{faber2016machine} and 13\% \cite{park2020developing}. Another key consideration for materials discovery is whether the model is able to generalise and make novel discoveries or simply interpolate current knowledge. Out of the 4,721 completed calculations 269 were assigned to AFLOW prototypes for which there were no isopointal prototypes in the training set. Of these 78 were confirmed to be below the convex hull of the union of the MP and WBM data sets. Developing workflows to directly target the discovery of structures for which no isopointal prototypes exist remains a key challenge for future work.}

\section{Discussion}

In this work, we introduced the concept of using coarse-graining to accelerate materials discovery.
We developed the framework of Wyckoff representation regression, \textit{Wren}, and applied it to predict the formation energy of materials.
\textit{Wren} collapses an infinite search space of atomic coordinates into a combinatorially enumerable search space enabling efficient data-driven exploration of materials space.
On a set of challenging tasks curated from the literature, we find that our approach can accurately map the phase diagrams of unseen chemical systems and is $\sim$3x better at finding stable materials than current methods based on elemental substitutions.

We developed a materials prospecting pipeline using \textit{Wren}. As a proof-of-concept, we identified 1,558 new materials below the known convex hull from just 5,675 calculations.
These results demonstrate that leveraging Wyckoff representation regression allows for more efficient and extensive expansion of computational material science databases.
Such efforts are crucial to expedite the search for a wide variety of industrially desirable materials required for the transition to a low-carbon economy, e.g. thermoelectrics \cite{wang2011assessing}, piezoelectrics \cite{tholander2016strong}, fast-ion conductors \cite{sendek2017holistic}, high voltage multi-valent cathode materials \cite{canepa2017odyssey}, and caloric materials \cite{zarkevich2017high}.

\section{Materials and Methods}

\subsection{\textit{Wren} Model Architecture}

The bulk of the \textit{Wren} architecture directly mimics that of \textit{Roost} \cite{goodall2020predicting} and we refer the readers there for an in-depth description of how the message passing is formulated.
The principle difference between the two architectures comes in that the nodes on the dense graph now represent the different Wyckoff positions rather than the different elemental species (See \autoref{fig:emb}).
The elemental information is encoded using the `matscholar' embedding from \cite{tshitoyan2019unsupervised} which has a dimensionality of $d_{el}=200$.
The remainder of the node embedding comprises the Wyckoff position embedding (described below) plus the fractional multiplicity of that Wyckoff position within the unit cell. The combined dimensionality of the Wyckoff proportion of the embedding is $d_{wyk}=445$.

To reduce the total dimension of the node embeddings, we project both the elemental and Wyckoff embeddings into lower-dimensional spaces using learnt affine transformations.
The low dimension embeddings are then concatenated to give the node embeddings.
In this work we chose $d_{el}^*=32$ and $d_{wyk}^*=32$ giving a total dimensionality of $d=64$ for the node embeddings.

We use 3 message passing layers, each with a single attention head.
We chose single-hidden-layer neural networks with 256 hidden units and LeakyReLU activation functions for both parts of the soft-attention mechanism.
The output network consists of a feed-forward neural network with skip connections and ReLU activation functions.
The output network used has 4 hidden layers containing 256, 256, 128, and 64 hidden units respectively.

\subsection{Wyckoff Position Embedding}
\begin{figure}
    \centering
    \includegraphics[width=0.35\textwidth]{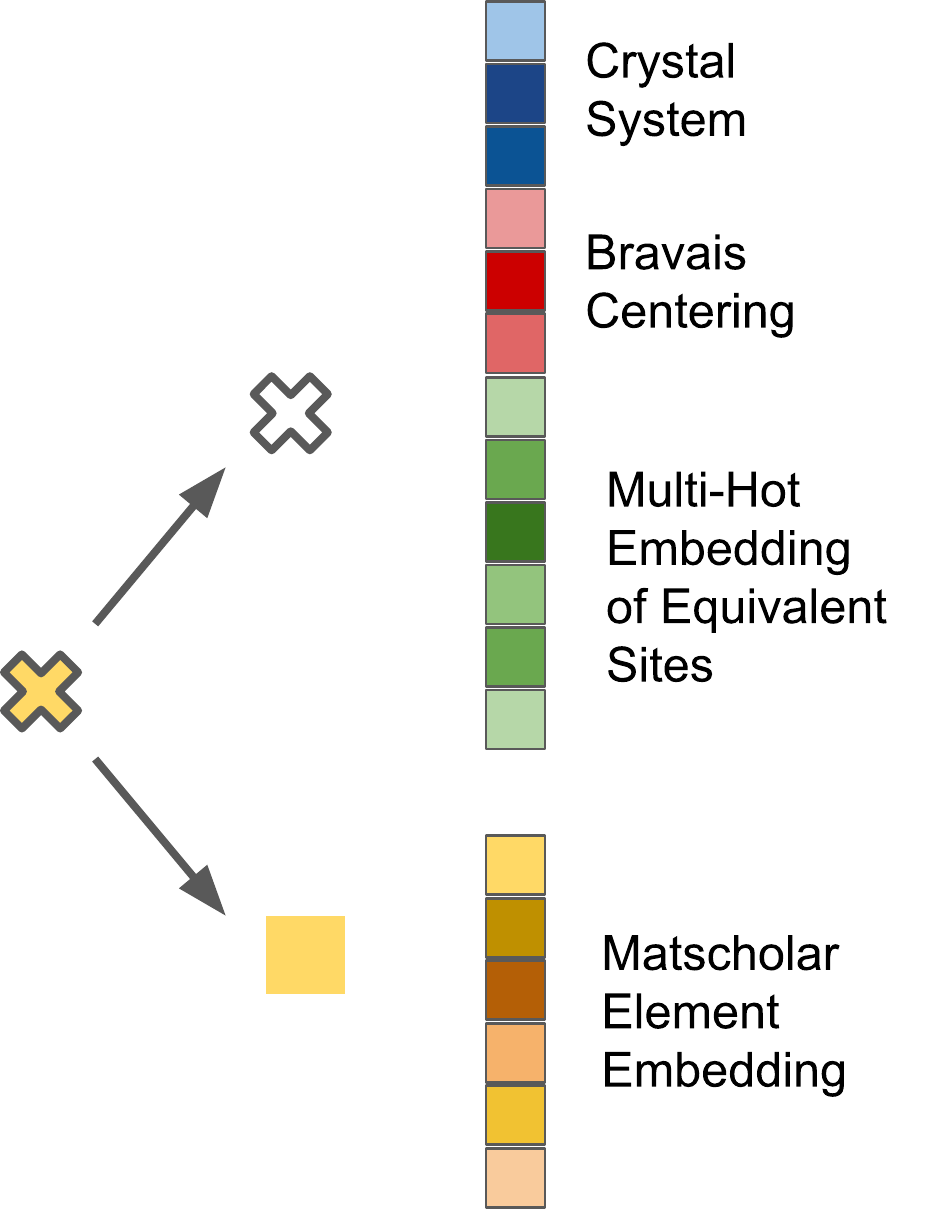}
    \caption{\textbf{Breakdown of different components of the Wyckoff position embeddings.}
    The Wyckoff position embeddings are made up of two parts. First, the Wyckoff proportion of the embedding that is comprised of 3 subsections encoding the crystal system, Bravais centring and equivalent sites in the Wyckoff positions. Second, the elemental embedding for which we take the `matscholar' embedding from \cite{tshitoyan2019unsupervised}.}
    \label{fig:emb}
\end{figure}

In total across the 230 crystallographic spacegroups in 3D there are 1,731 different Wyckoff positions.
The embedding we use is made up of 3 parts; a one-hot encoding of the crystal system (of which there are 6), a one-hot encoding of the Bravais lattice centring (of which there are 5), and an encoding constructed from the sum of multi-hot encodings of the equivalent sites within a given Wyckoff position (See \autoref{fig:emb}).
To construct the multi-hot encodings, we first collate all the sites within all the allowed Wyckoff positions as recorded on the Bilbao crystallographic server \cite{aroyo2006bilbao}.
Each site can be broken apart into its offset and algebraic terms (whether the position corresponds to a line, a plane, etc.), e.g.
\begin{align}
    (-x+y+1/4, y, z+3/4) = \ &(1/4, 0, 3/4) \ + \nonumber \\
    &(-x+y, y, z)
\end{align}
From this, we construct separate one-hot encodings for the unique algebraic and unique offset positions.
We end up with 185 unique algebraic positions and 248 unique offset positions.
A Wyckoff position is then represented by a sum of the embeddings of all the allowed sites.
The resulting embedding has a dimensionality of 444 with the 1,731 Wyckoff positions being encoded into 1038 unique embeddings. 
This embedding is designed to try and expose as many possible correlations as possible that might exist between different Wyckoff positions. As an illustrative example the embeddings for the f Wyckoff position of $Fm\Bar{3}$ (No. 202), $F432$ (No. 209), and $Fm\Bar{3}m$ (No. 225) are all the same. This arises as they are all face-centred cubic lattices that describe the positions of 32 atoms within the unit cell at ((0, 0, 0), (0, 1/2, 1/2), (1/2, 0, 1/2), (1/2, 1/2, 0)) $\oplus$ ((x, x, x), (-x, -x, x), (-x, x, -x), (x, -x, -x), (-x, -x, -x), (x, x, -x), (x, -x, x), (-x, x, x)), where x is a free-coordinate of the Wyckoff positions. The embedding vector has 4's in the positions corresponding to the (x, x, x), (-x, -x, x), (-x, x, -x), (x, -x, -x), (-x, -x, -x), (x, x, -x), (x, -x, x), (-x, x, x) algebraic terms and 8's in the positions corresponding to the (0, 0, 0), (0, 1/2, 1/2), (1/2, 0, 1/2), (1/2, 1/2, 0) offset terms. In principle further engineering of this embedding could be carried out to encode more prior knowledge, however, for the sizes of data set considered in this work the benefit of doing so is likely to be marginal.

\subsection{Invariance to Equivalent Wyckoff Representations}

\begin{figure}
    \centering
    \includegraphics[width=0.30625\textwidth]{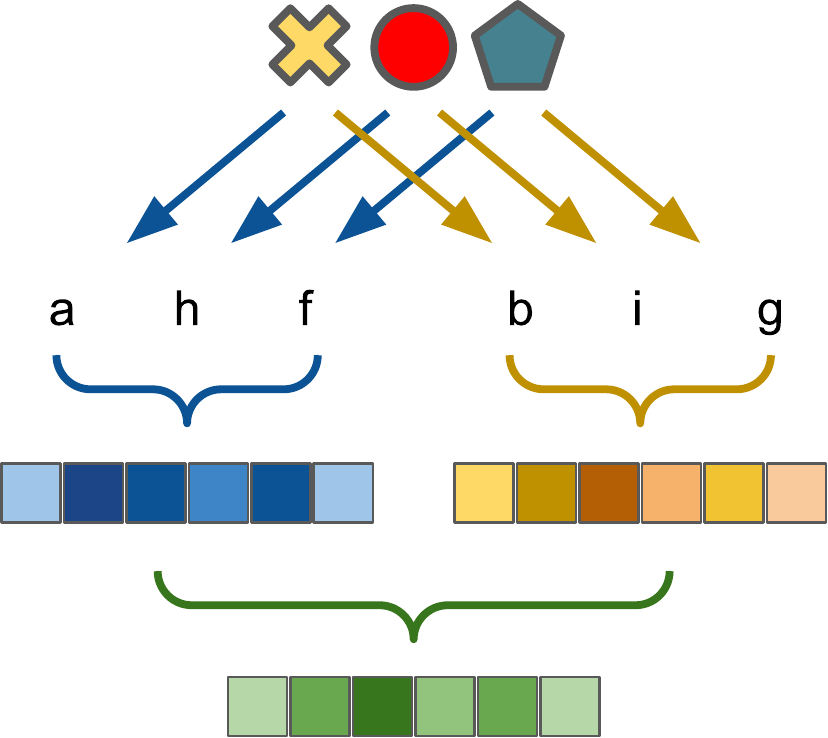}
    \caption{\textbf{On-the-fly augmentation of equivalent Wyckoff representations ensures invariance to equivalent descriptions.}
    The labelling of Wyckoff positions includes a choice of setting, in order to ensure that our model is invariant to these choices we perform on the fly augmentation of all equivalent Wyckoff representations and then average the augmented embeddings before they are fed into the output network.}
    \label{fig:aug}
\end{figure}

The categorisation of Wyckoff positions depends on a choice of origin \cite{boyle1973origin}.
As such, there is not a unique mapping between the crystal structure and the Wyckoff representation.
To ensure the model is invariant to the choice of origin, we perform on-the-fly augmentation of Wyckoff positions with respect to this choice of origin (See \autoref{fig:aug}).
The augmented representations are averaged at the end of the message passing stage to provide a single representation of equivalent Wyckoff representations to the output network.
By pooling at this point we ensure that the model is invariant and that its training is not biased towards materials for which many equivalent Wyckoff representations exist.

\subsection{Evaluation of Wyckoff Positions}

For this work, we primarily make use of \texttt{spglib} \cite{togo2018texttt} to determine the spacegroup and Wyckoff positions for the structures in the data sets.
We set the tolerance thresholds as 0.1 {\AA} for positions and $5^{\circ}$ for angles (Note, these are the same tolerances as used in Materials Project to calculate the spacegroup).
In real materials, we often observe some degree of off-site relaxation away from high-symmetry sites. Depending on the level of anisotropy and the symmetry finder's tolerance threshold, this might result in materials being classed as $P1$. As lower symmetry Wyckoff representations encode less information about the structure, the symmetry finder tolerance is an important hyper-parameter to bear in mind.
However, preliminary investigations showed that varying the tolerance threshold between typical values of 0.01 {\AA} and  0.1 {\AA} did not significantly impact model accuracy on the TAATA data set.
We note that as an alternative to manual selection of tolerance hyper-parameters, symmetry finders with adaptive tolerances, such as \texttt{aflow-sym} \cite{hicks2018aflow}, could be used for the identification of the spacegroup and Wyckoff positions. However, given that we did not observe any appreciable improvement in accuracy using \texttt{aflow-sym} and adaptive-schemes are typically associated with greater computational cost, \texttt{spglib} was picked over other symmetry finders due to its speed.

\subsection{Model Training}
Throughout this work, we train Deep Ensembles of 10 models starting from different random initialisations for each data setup and architecture considered.

All the models examined in this work were trained using the \textit{AdamW} optimiser \cite{loshchilov2019adamw} with a fixed learning rate of $3\times 10^{-4}$.
A mini-batch size of 128 and a weight decay parameter of $10^{-6}$ were used for all the experiments.
The models were trained for 400 epochs (cycles through the training set).

Formally Deep Ensembles require the use of a proper scoring rule for training.
Therefore, we train all models to minimise the following robust L1 loss function which is an example of a proper scoring rule for regression \cite{nix1994estimating, kendall2017uncertainties},
\begin{equation}
\label{eq:loss}
   \mathcal{L} = \sum_i \dfrac{\sqrt{2}}{\hat{\sigma}_{a,\theta}(x_i)} \ \|y_i - \hat{\mu}_\theta(x_i)\|_1 + \text{log}\Big(\hat{\sigma}_{a,\theta}(x_i)\Big)
\end{equation}
where $\hat{\mu}_\theta(x_i)$ and  $\hat{\sigma}_{a,\theta}(x_i)^2$ are a predictive mean and predictive aleatoric variance outputted by the model, and $y_i$ is the target label.

The expectations, $\hat{y}(x_i)$, and epistemic uncertainties, $\hat{\sigma}_{a}(x_i)^2$ from the ensemble are calculated as
\begin{equation}
    \hat{y}(x_i) = \frac{1}{W} \sum_w^W  \hat{\mu}_{\theta_w}(x_i),
\end{equation}
\begin{equation}
    \hat{\sigma}_{e}^2(x_i) = \frac{1}{W} \sum_w^W \Big(\hat{y}(x_i) -\hat{\mu}_{\theta_w}(x_i)\Big)^2,
\end{equation}
where the index $w$ runs over the $W$ members of the ensemble. The total uncertainty of the ensemble expectation is simply the sum of the epistemic contribution and the average of the aleatoric contributions from each model in the ensemble.
\begin{equation}
    \hat{\sigma}^2(x_i) = \hat{\sigma}_{e}^2(x_i) + \frac{1}{W} \sum_w^W \hat{\sigma}_{a,\theta_w}^2(x_i).
\end{equation}

\subsection{Canonicalisation and Cleaning}

\begin{table}

    \centering
    \caption{Table showing the the impact of cleaning criteria on the MP data set.}
    \label{tab:clean-mp}
    {\color{black}\begin{tabular}{lc}
    \hline
    Filter & Number \\
   \hline
    Full Data Set & 139,367 \\
    Lowest Energy Canonical Representations & 129,190 \\
    Formation Energy less than 5 eV per atom & 129,176 \\
    Less than 16 Wyckoff Positions & 108,656 \\
    Less than 64 sites in Crystal Structure & 105,057 \\
    Volume per site less than 500 {\AA}$^3$ & 104,878 \\
    \hline
    \end{tabular}}

    \caption{Table showing the the impact of cleaning criteria on the WBM data set.}
    \label{tab:clean-wbm}
    {\color{black}\begin{tabular}{lc}
    \hline
    Filter & Number \\
   \hline
    Full Data Set & 257,486 \\
    Lowest Energy Canonical Representations & 224,498 \\
    After Removal of Duplicates Found in MP & 217,085 \\
    Excluding Pure Systems & 216,877 \\
    Formation Energy less than 5 eV per atom & 216,859 \\
    Less than 16 Wyckoff Positions & 216,819 \\
    Less than 64 sites in Crystal Structure & 216,807 \\
    Volume per site less than 500 {\AA}$^3$ & 216,806 \\
    \hline
    \end{tabular}}

    \caption{Table showing the the impact of cleaning criteria on the TAATA data set.}
    \label{tab:clean-taata}
    {\color{black}\begin{tabular}{lc}
    \hline
    Filter & Number \\
   \hline
    Full Data Set & 12,815 \\
    Lowest Energy Canonical Representations & 9,688 \\
    Less than 16 Wyckoff Positions & 9,490 \\
    Less than 64 sites in Crystal Structure & 9,190 \\
    Volume per site less than 500 {\AA}$^3$ & 9,190 \\
    \hline
    \end{tabular}}

\end{table}

All the data used to train models in this work went through a canonicalisation and cleaning process.
Tables \ref{tab:clean-mp}, \ref{tab:clean-wbm}, and \ref{tab:clean-taata} show how much data is discarded at each stage.

The canonicalisation stage removes higher energy structures that have equivalent Wyckoff representations to other structures in the data set.

We adopt the same canonicalisation scheme as used by the AFLOW prototype encyclopedia \cite{mehl2017aflow, hicks2019aflow}. Most structures removed by canonicalisation are triclinic, as the lack of symmetries in triclinic systems results in many distinct structures mapping to the same Wyckoff representation.

\textcolor{black}{For the WBM data set we carried out cleaning based on the relaxed structures and relaxed Wyckoff representations. We removed elemental structures in the WBM data set to ensure that our end-points for calculating formation energies were consistent.
The union of the MP and WBM data sets used to train the \textit{Wren} model for the prospective validation made use of an earlier cannonicalisation scheme which led to it containing some duplicated Wyckoff representations. In total the union of MP and WBM used contained 322,915 materials.}

\subsection{Prospective Density Functional Theory Settings}

The validation of predictions in our materials prospecting pipeline was carried out using Kohn-Sham DFT with the plane wave pseudopotential code \texttt{VASP} \cite{kresse1996efficiency, kresse1996efficient}.
Projector augmented wave (PAW) type pseudopotentials \cite{blochl1994projector, kresse1999ultrasoft} were used with the Perdew-Burke-Ernzerhof (PBE) generalised gradient approximation exchange-correlation functional \cite{perdew1996generalized}.
All calculations were done using a 520 eV plane-wave energy cutoff. The pseudopotentials and Hubbard-$U$ values were selected to ensure compatibility with data contained in the Materials Project.
The Materials Project \texttt{MaterialsProjectCompatibility} correction scheme implemented in \texttt{pymatgen} was applied to allow the mixing of GGA and GGA+U calculations \cite{jain2011formation}.
We used the High-Throughput Toolkit (\texttt{httk} v1.0) introduced in Ref. \cite{armiento2020httk} to manage the calculations.

\section*{Data Availability}
The relaxed and initial structures for the TAATA data set \textcolor{black}{are available from \href{https://doi.org/10.5281/zenodo.5530535}{https://doi.org/10.5281/zenodo.5530535}}.

The MP data set used for this work was queried from \href{https://materialsproject.org}{https://materialsproject.org} \cite{jain2013commentary} using the Materials API \cite{ong2015materials}. The data was queried from Supplemental Database Release V2021.03.22.

The relaxed structures for WBM data set used in this work are available from \href{https://archive.materialscloud.org/record/2021.68}{https://archive.materialscloud.org/record/2021.68}. The pre-relaxation structures were obtained from the original authors on request.

The ICSD \cite{bergerhoff1987crystallographic, hellenbrandt2004inorganic} can be accessed under license from FIZ Karlsruhe—Leibniz Institut for Information Infrastructure at \href{https://icsd.products.fiz-karlsruhe.de/}{https://icsd.products.fiz-karlsruhe.de/}.

\textcolor{black}{The relaxed and initial structures for all completed prospective calculations are available from \href{https://doi.org/10.5281/zenodo.6345276}{https://doi.org/10.5281/zenodo.6345276}.}

\section*{Code Availability}

A working version of the code is available at \href{https://github.com/CompRhys/aviary}{https://github.com/CompRhys/aviary}.

\section*{Acknowledgements}
REAG and AAL acknowledge the support of the Winton Programme for the Physics of Sustainability. AAL acknowledges support from the Royal Society. FAF acknowledges funding from the Swiss National Science Foundation (Grant No. P2BSP2\_191736). RA and ASP acknowledge support from the Swedish Research Council (VR) Grant No. 2020-05402  and the Swedish e-Science Centre (SeRC). The computations were enabled by resources provided by the Swedish National Infrastructure for Computing (SNIC) at NSC partially funded by the Swedish Research Council through grant agreement No. 2018-05973.

The authors thank Janosh Riebesell for valuable discussions regarding this work and Hai-Chen Wang and co-authors for providing the pre-relaxation structures for the WBM data set.

\section*{Author Contributions}
REAG: Methodology, Software, Investigation, Visualisation, Writing - Original Draft. ASP: Methodology, Investigation, Validation. FAF: Conceptualisation, Supervision, Writing - Review \& Editing. RA: Conceptualisation, Supervision, Writing - Review \& Editing. AAL: Supervision, Writing - Review \& Editing.


\section*{Competing Interests}
The authors declare no competing interests.

\newpage

\bibliographystyle{unsrt}
\bibliography{references.bib}

\end{document}


\title{SI: Rapid Discovery of Stable Materials by Coordinate-free Coarse Graining}

\author{Rhys E. A. Goodall}
    \affiliation{Department of Physics, University of Cambridge, United Kingdom}

\author{Abhijith S. Parackal}
    \affiliation{Department of Physics, Chemistry and Biology, Linköping University, Sweden}

\author{Felix A. Faber}
    \affiliation{Department of Physics, University of Cambridge, United Kingdom}

\author{Rickard Armiento}
    \affiliation{Department of Physics, Chemistry and Biology, Linköping University, Sweden}

\author{Alpha A. Lee}
    \affiliation{Department of Physics, University of Cambridge, United Kingdom}


\keywords{Materials Science, Machine Learning, Representation Learning}

\maketitle

\section{Symmetry Changes During Symmetry Constrained Relaxations}
Carrying out a symmetry constrained relaxation in \texttt{VASP} prevents symmetry breaking, but in some instances, we can see the merger of distinct Wyckoff positions into higher symmetry Wyckoff positions.
This can lead to a change in the Wyckoff representation in screening workflows between the pre-relation and relaxed structures.
To ensure that we have a well-defined map, we train \textit{Wren} models to learn the target properties as a function of the relaxed Wyckoff representation.

A change in Wyckoff representation was observed in \textcolor{black}{29,417 out of 257,486 materials in the WBM data set}.
Consequently, the Wyckoff representation appears to change during relaxation for approximately 10\% of candidates produced using prototype-based substitution workflows.

\begin{figure}[b]
    \centering
    \includegraphics[width=0.49\textwidth]{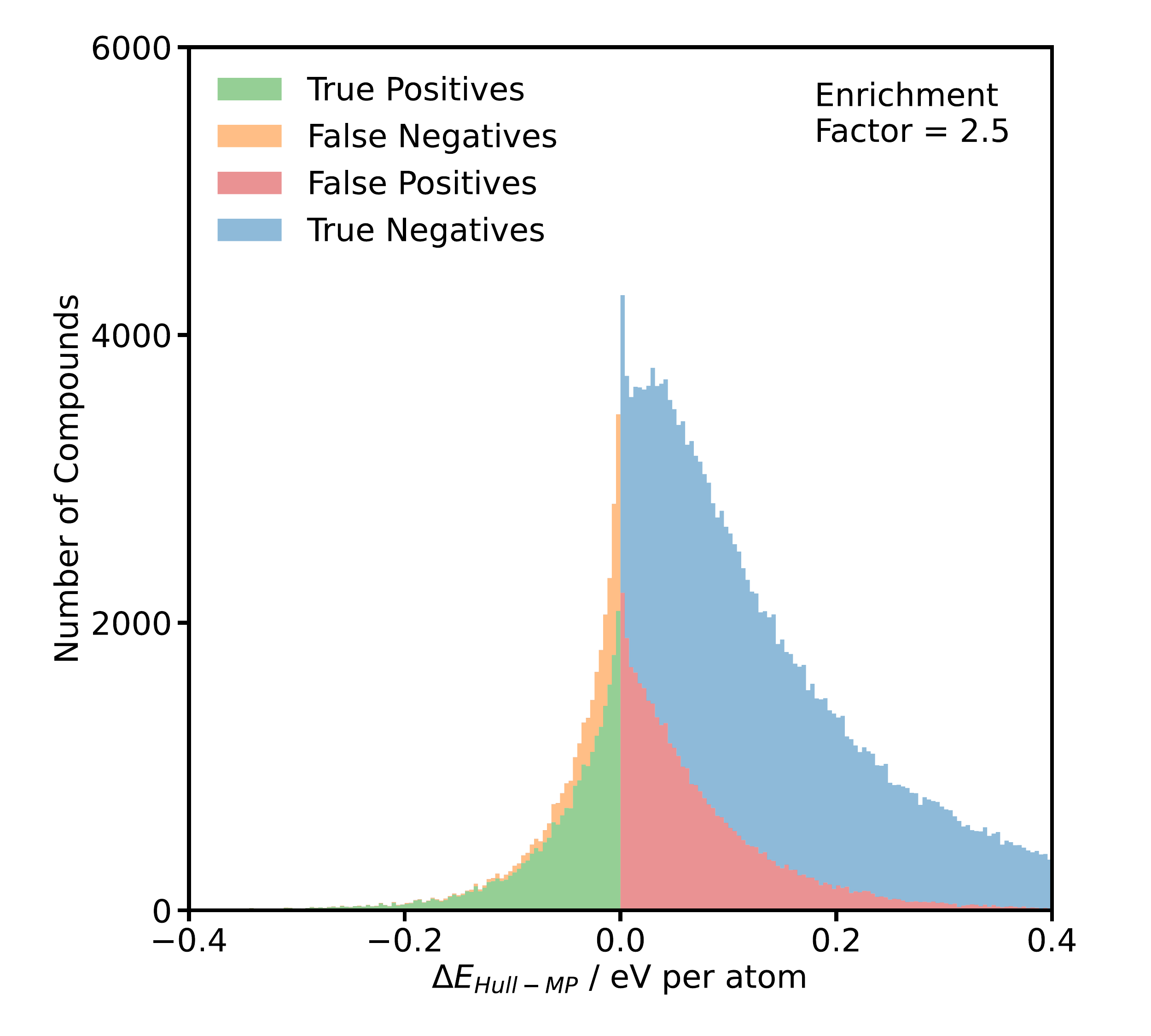}
    \caption{\textbf{\textit{Wren} efficiently identifies stable structures when screening diverse chemical spaces.}
    Histogram of the energy to the convex hull for materials in the WBM data set.
    The histogram is broken down into true positives, false negatives, false positives, and true negatives based on whether the \textit{Wren} model predicts candidates to be below the known convex hull.
    \textit{Wren} exhibits a high recall with the majority of materials below the convex hull being correctly identified by the model.
    }
    \label{fig:quasi-stable}
\end{figure}

\begin{figure}[b]
    \centering
    \includegraphics[width=0.49\textwidth]{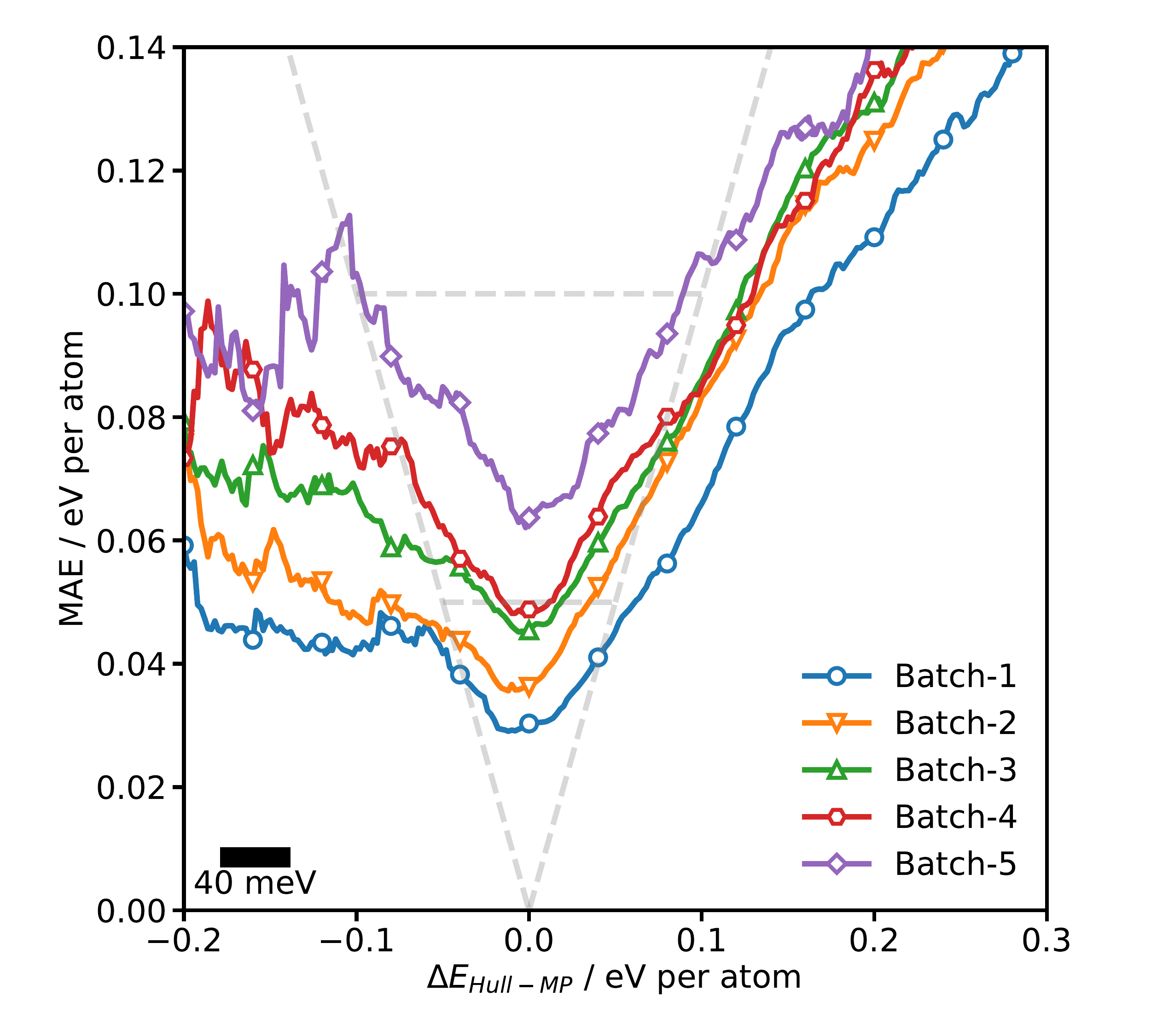}
    \caption{\textbf{Rolling MAE of the \textit{Wren} model on the WBM data set batches.}
    Rolling mean absolute error of \textit{Wren} on the batches of the WBM data set as the energy to the convex hull is varied.
    We take the pre-relaxation WBM Wyckoff representation as input.
    A windowing period of 40 meV per atom is used when calculating the rolling average.
    The curves show that later batches, believed to be more likely to be chemically dissimilar to the training data, incur higher average errors.
    }
    \label{fig:rolling-batches}
\end{figure}

\section{Enrichment as a Function of Distance to the Convex Hull}

\autoref{fig:quasi-stable} shows a stacked histogram of the energy to the convex hull of the full MP data set for the materials from the WBM data set.
The histogram sections are coloured according to whether the \textit{Wren} model correctly predicts a candidate to be below the known convex hull given its Wyckoff representation.

\section{Extrapolation in Material Space}

The WBM data set was generated using an iterative workflow where successful candidates from the first batch were included when generating candidates for the second batch.
As a consequence, candidates considered in later batches are likely to be less similar to materials contained in the MP data set.
We use this stratification to probe how \textit{Wren}'s performance changes as it is asked to make larger and larger extrapolations.
We can make use of this stratification to probe how \textit{Wren}'s performance changes as it is asked to make larger and larger extrapolations by looking at how the mean absolute error changes as a function of the distance from the convex hull for the different batches.
In \autoref{fig:rolling-batches} we see that, whilst the overall shape of the curves remains the same as the V-shape seen for the full data set, there is an offset between the batches with later batches incurring higher errors on average.

\begin{figure*}[!th]
    \centering
    \includegraphics[width=0.98\textwidth]{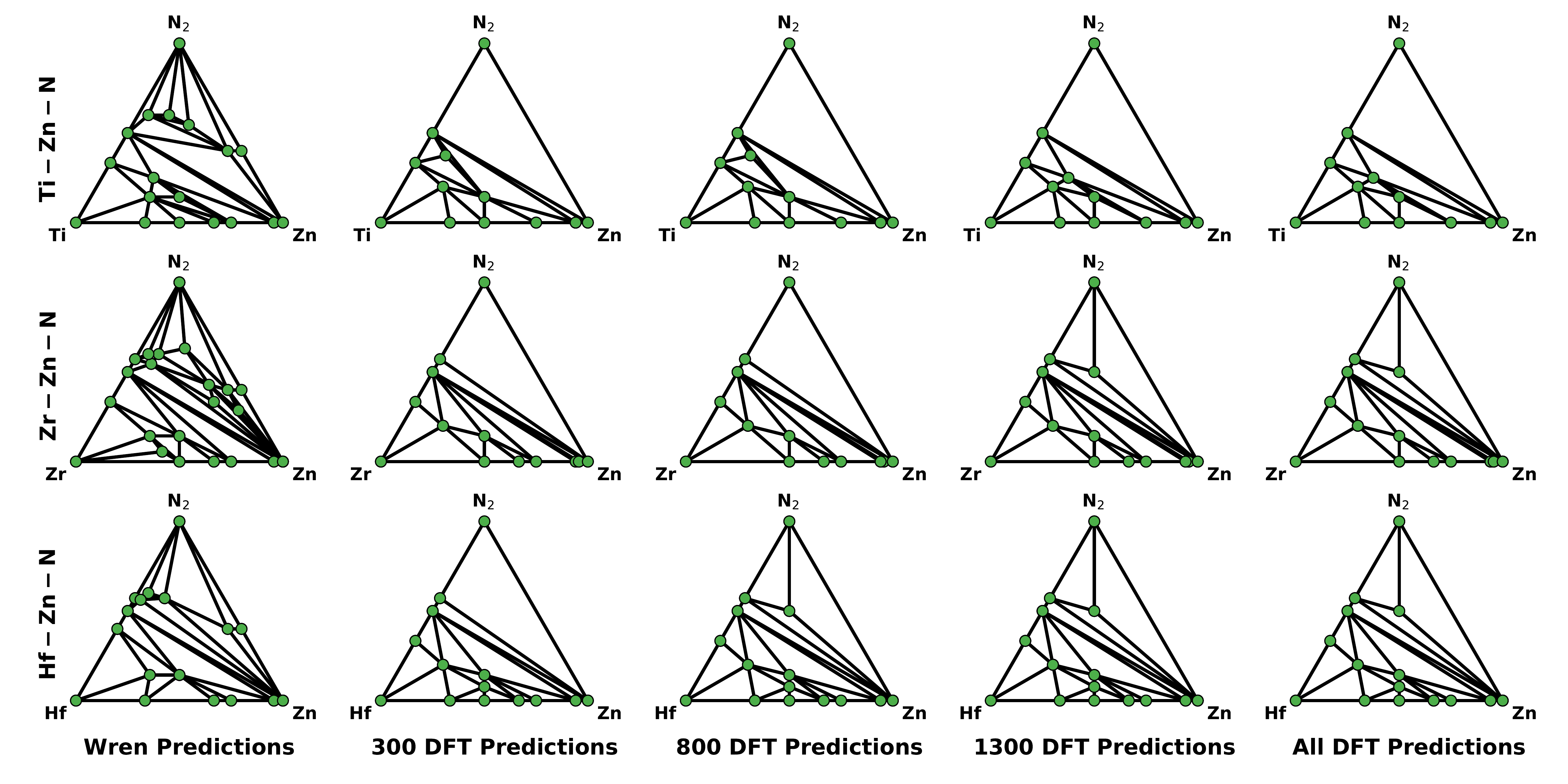}
    \caption{
    \textbf{Phase Diagrams for the TAATA Data Set as \textit{Wren} is used to Triage Calculations.}
    \textcolor{black}{Whilst the zero-shot phase diagrams generated by the \textit{Wren} model do not appear visually similar to the DFT computed phase diagram, we see that using the energy above the hull to the zero-shot convex hull is an effective surrogate for stability. Using this prediction to triage candidates allows for phase diagrams that closely resemble the DFT computed phase diagrams to be constructed with as little as 300 calculations.}
    }
    \label{fig:taata-pds}
\end{figure*}

\textcolor{black}{
\section{TAATA Data Set Details}
The TAATA data set was orignally constructed in \cite{tholander2016strong} using an exhaustive prototyping strategy.
Initial prototypes were extracted from the Inorganic Crystal Structure Database (ICSD) and the Crystallography Open Database (COD).
All inequivalent structures with one anion and two cations containing less that 41 atoms were used.
In total, 2,444 structures satisfied these criteria and were used for prototyping.
In each case the anion was replaced by N, and the cations were substituted with one of the \{Ti, Zr, Hf\} species and Zn.
Both permutations of \{Ti, Zr, Hf\} species and Zn were considered.
The starting volume of each structure was rescaled to give the same atom to unit cell ratio as in the ground state structures.
In addition to ternary, binary, and elemental structures from the Materials Project database were also included in the search.}

\section{Phase Diagram Construction}

\textcolor{black}{\autoref{fig:taata-pds} shows phase diagrams constructed from the initial Wren predictions as well as the nominal ground truth phase diagram when all the DFT calculated energies are considered. In order to demonstrate that \textit{Wren} aids our exploration of unexplored systems we also show the resulting phase diagrams after 300, 800 and 1,300 calculations have been carried out after triaging calculations using the predicted energy above the convex hull constructed using the predictions of the \textit{Wren} model. These intermediary plots show that whilst superficially the zero-shot hulls generated by \textit{Wren} do not match the DFT computed hulls using the \textit{Wren} predictions to triage calculations gives phase diagrams that closely resemble the exhaustive DFT computed phase diagram, each missing just a single stable phase, after just 300 calculations.}

\textcolor{black}{In the manuscript we choose to look at how well the approach performs on selecting candidates within 10, 20 and 30 meV per atom of the DFT computed convex hull in order to collect more reliable statistics. If we focus on just the DFT computed stable phases, we see that our model does rank a stable phase outside the top 300 in each of the systems studied. Checking these erroneous predictions reveals that they are all structures for which no isopointal analogues exist in the training data. It is worth noting that the fact that TAATA considers three group 4 elements, \{Ti, Zr, Hf\}, makes the prediction task significantly more challenging as we have no prior tertiary examples that share the same group chemistry.}

\begin{figure}
    \centering
    \includegraphics[width=0.49\textwidth]{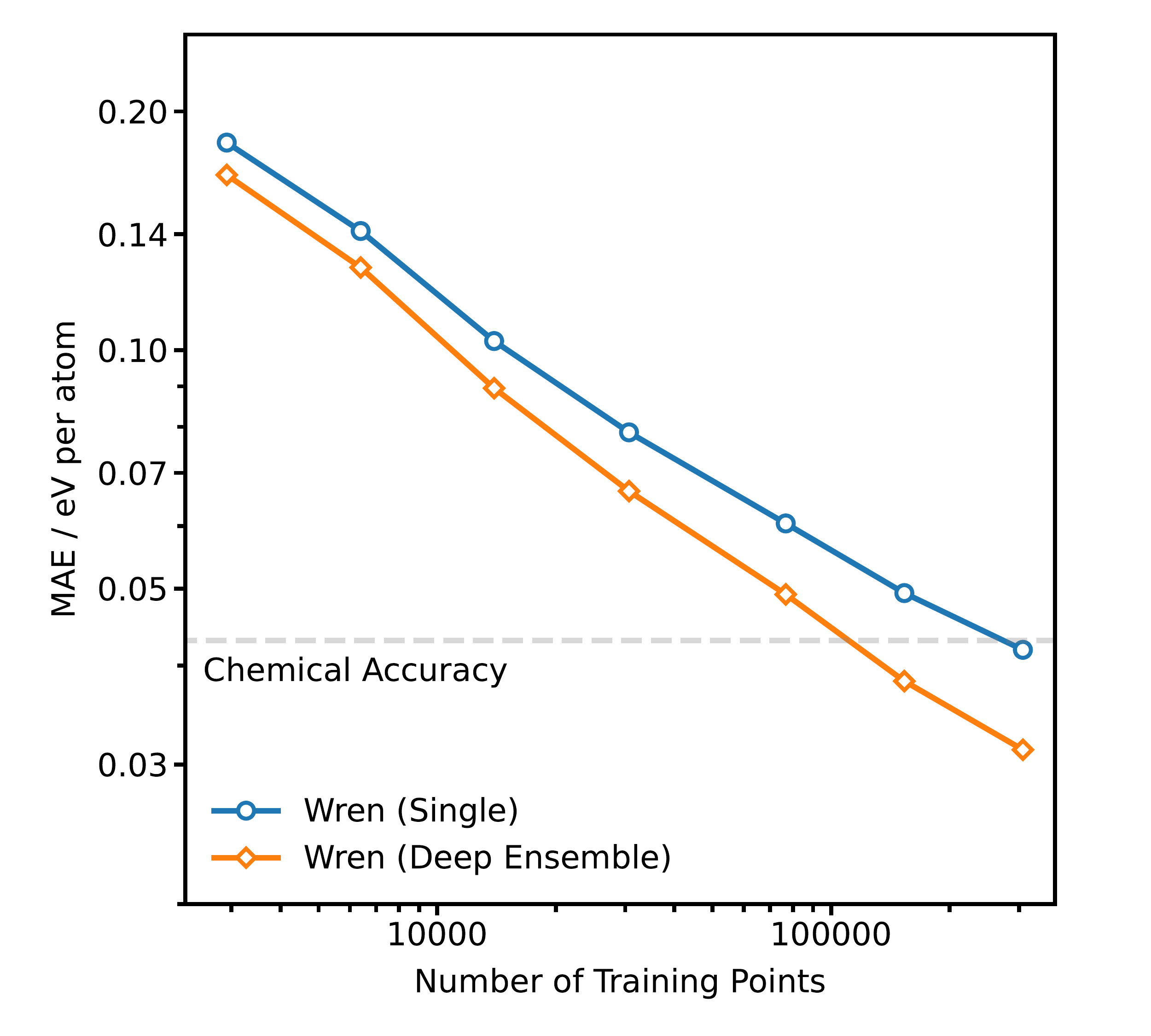}
    \caption{
    \textbf{Learning curve for the \textit{Wren} model with increasing training data.}
    Learning curves for both single and Deep Ensembles of \textit{Wren} models \textcolor{black}{shown on log-log scales}.
    The curves are produced by plotting the error on a fixed test set (here 5\% of the union of the MP and WBM data sets) as the amount of data used to train the model is increased.
    A grey guideline shows the size of the test set.
    The learning curve shows that a power-law relationship exists between the amount of training data and the MAE of the trained model.
    }
    \label{fig:learning-curve}
\end{figure}

\section{Learning Efficiency}

Applications in material science, particularly the investigation of functional properties of materials, often encounter issues with data scarcity.
Given this, the learning efficiencies of models used in material science applications are of critical importance.
The best way to probe this is through the construction of learning curves \cite{muller1996numerical, faber2018alchemical}.
A learning curve depicts the error of a model on a fixed test set as the amount of data used to train the model is varied.
We see in \autoref{fig:learning-curve} that both individual \textit{Wren} models and Deep Ensembles of \textit{Wren} models follow unbroken power law relationships.
Consequently, we would expect that as more data is obtained the accuracy of the proposed model should continue to improve.

\begin{figure}
    \centering
    \includegraphics[width=0.49\textwidth]{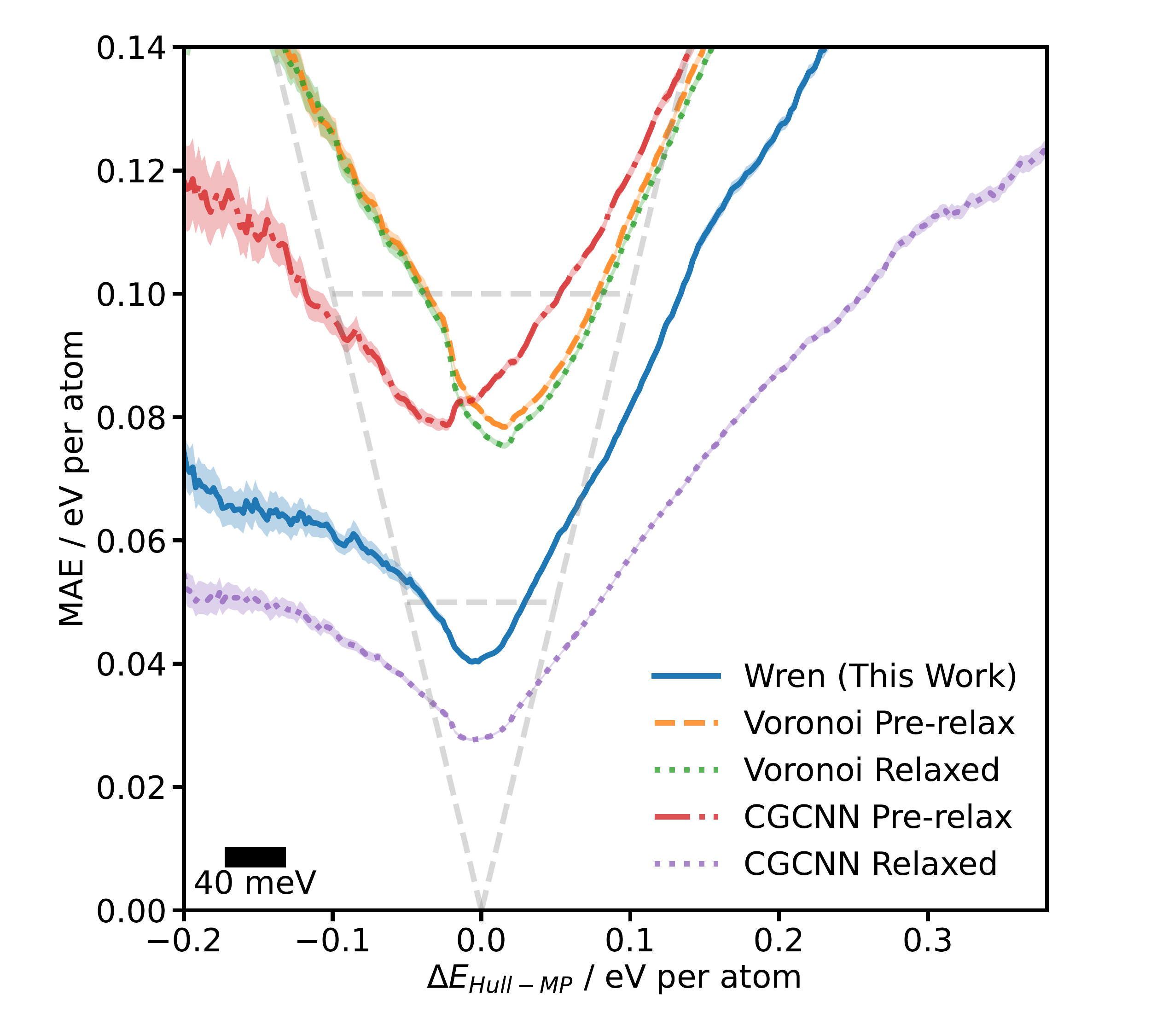}
    \caption{\textbf{Comparison of Errors of \textit{Wren}, \textit{Voronoi}, and \textit{CGCNN} Models on WBM with Distance to the MP Convex Hull.}
    Figure replicates experimental setup of Figure 2 comparing the \textit{Voronoi} model against \textit{CGCNN} and \textit{Wren}.
    A scale bar is shown for the windowing period of 40 meV per atom used when calculating the rolling average.
    Grey guidelines highlight 100 meV per atom, 50 meV per atom and the MAE = $|\Delta E_{\text{Hull-MP}}|$.
    The \textit{Wren} model is more accurate than both the \textit{Voronoi} model and \textit{CGCNN} on pre-relaxation structures but, as expected, it is less accurate than \textit{CGCNN} on relaxed structures.
    \textit{Wren} is more accurate than the \textit{Voronoi} model on relaxed structures highlighting the strengths of \textit{Wren} and the Wyckoff representation.}
    \label{fig:rolling-compare}
\end{figure}

\section{Alternative Structure-based Models for Materials Discovery}

To contextualise the performance of the \textit{Wren} model, we compare against \textit{CGCNN}, a well-established message passing neural network for predicting the properties of inorganic materials based on the ``crystal-graph'' of local atomic environments \cite{xie2018crystal}, and a baseline consisting of a Random Forest \cite{breiman2001random} paired with the Voronoi crystal structure attributes \cite{ward2017including} plus \textit{Magpie} composition descriptors \cite{ward2016general} as proposed in \cite{ward2017including} that we refer to as the \textit{Voronoi} model.

In prototype-based substitution workflows, such as that used to generate the WBM data set \cite{wang2021predicting}, the distributions of bond distances seen before relaxation can be highly unphysical as the original species often have very different atomic radii to the substituted species.
Consequently, models such as \textit{CGCNN} that directly encode local environments based on bond distances are likely to suffer from significant degradation in performance when moving from screening relaxed crystal structures to screening pre-relaxation crystal structures \cite{park2020developing}.
The Voronoi crystal structure attributes are constructed conceptually distinct manner, using the areas of the Voronoi facets around a given site to weigh local property differences between sites in such a way that the resulting features are invariant to changes in volume.
This is a compelling property as one of the principal changes that occurs when relaxing a prototype structure after chemical substitution is the change in the unit cell volume.

As expected, we see that using \textit{CGCNN} to predict the stability of candidate materials based on their relaxed structures is highly accurate.
However, as previously noted, in prospective workflows we do not have access to relaxed crystal structures.
When we compare the stability predictions obtained when using \textit{CGCNN} to estimate the stability of candidate materials based on their pre-relaxation structures, we see a substantial degradation in performance in keeping with the results of prior works \cite{park2020developing}.

Turning to the \textit{Voronoi} model, we see that robustness of the Voronoi crystal structure attributes is confirmed in the observation that the relaxed and pre-relaxation structures make similar predictions on the WBM data set after having been trained on the MP data set (see \autoref{fig:rolling-compare}).
Quantitatively a mean absolute deviation of 19 meV per atom between the pre-relaxation and relaxed structure predictions is seen for the \textit{Voronoi} model with is small compared to the MAEs of \textcolor{black}{145 and 142} eV per atom respectively.
Despite this robustness, the \textit{Voronoi} model results in a much larger error than \textit{Wren}.

\begin{figure}[t]
    \centering
    \includegraphics[width=0.49\textwidth]{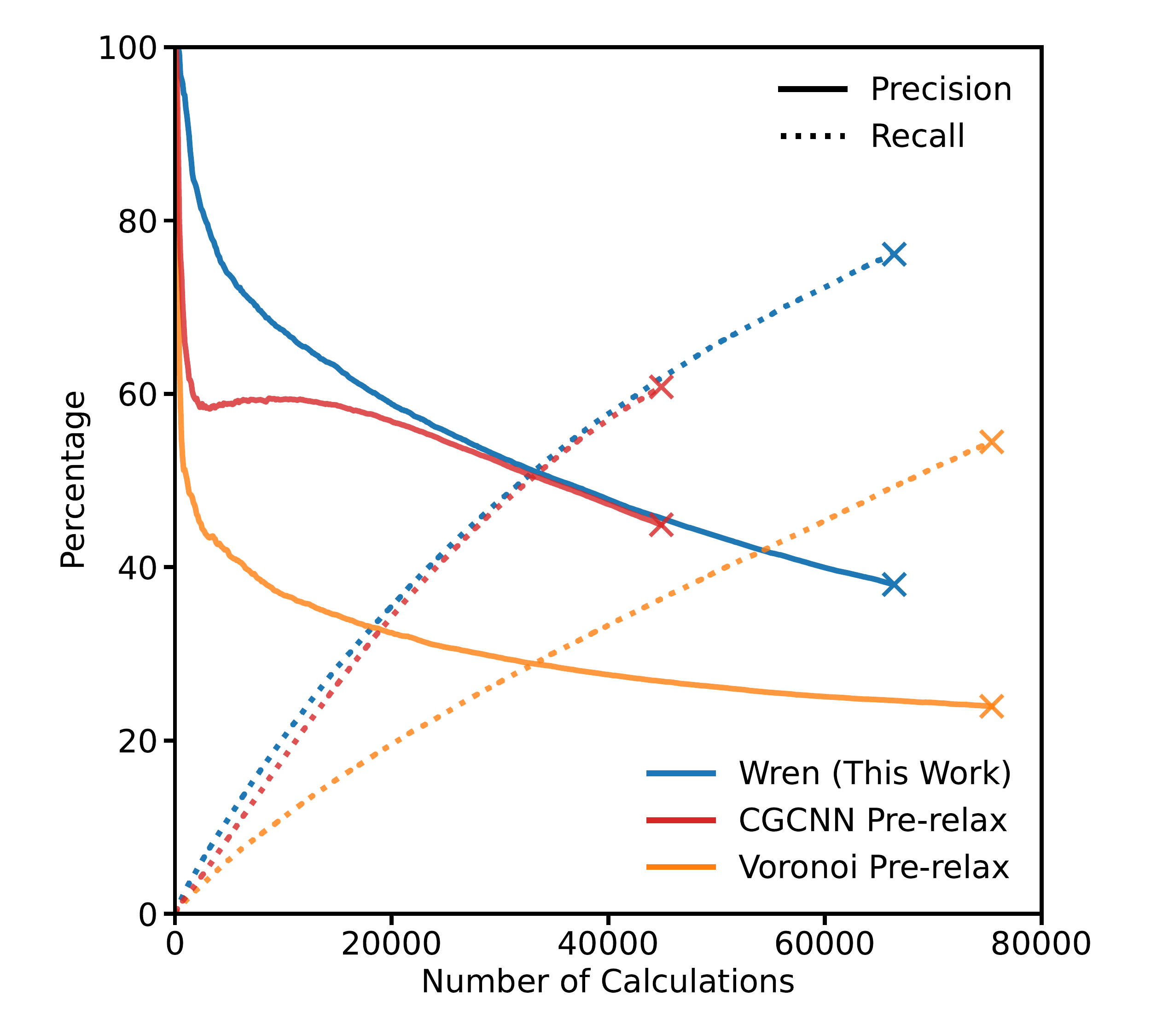}
    \caption{\textbf{Variation in Precision and Recall for \textit{Wren}, \textit{Voronoi}, and \textit{CGCNN} Models with Calculation Count.}
    Precision and Recall on the WBM data set for \textit{Wren}, \textit{Voronoi}, and \textit{CGCNN} as the number of calculations increased.
    The precision is shown with a dashed line and the recall is shown with a dotted line.
    Crosses are used to mark the termination points where the models no longer believe there are anymore materials in the WBM data set below the MP convex hull.
    Initially both \textit{Voronoi} and \textit{CGCNN} models have much lower precision than \textit{Wren}.
    However the precision of \textit{CGCNN} plateaus and then is upper-bounded by that of the \textit{Wren} model until it terminates with a recall of \textcolor{black}{61}\%.
    The \textit{Voronoi} model has much lower precision and terminates with a recall of \textcolor{black}{55}\% after $\sim$ 77,000 calculations.
    }
    \label{fig:prec-rec-wbm}
\end{figure}

We note that even though screening for stability using the pre-relaxation structures is not very accurate, using structure-based models in this manner still leads to enrichment in virtual screening workflows, for example, Park et al. have used this approach alongside their ``improved''-\textit{CGCNN} model \cite{park2020developing} to accelerate the discovery of novel stable compounds within the ThCr\textsubscript{2}Si\textsubscript{2} structure-type achieving a precision of 13\%.
Whilst such workflows offer an improvement over conventional screening workflows in terms of enrichment, the limited accuracy of using structure-based models in this manner leads to low recall -- the \textit{Voronoi} model achieves a recall of \textcolor{black}{55\%} and the \textit{CGCNN} model achieves a recall of \textcolor{black}{61\%} of the materials in the WBM data set below the known convex hull compared to a recall of \textcolor{black}{76\%} for \textit{Wren} (see \autoref{fig:prec-rec-wbm}).

Accordingly, several attempts have been made to improve the performance of structure-based models when making stability predictions on pre-relaxation structures but attempting to make models that are robust to changes that occur during structure relaxation.
The symmetry-labelled Voronoi graph convolutional neural network of \cite{jorgensen2019materials} tries to do this by discarding explicit distance information.
Instead, edges in the graph are labelled by the approximate symmetry of the Voronoi facets.
This leads to a smaller deterioration in accuracy than seen for distance-based models.
More recently, the \textit{BOWSR} algorithm \cite{zuo2021accelerating} has been introduced to directly minimise the domain shift between pre-relaxation and relaxed structures using a surrogate model.
The algorithm uses Bayesian optimisation to adjust the free parameters of pre-relaxation symmetrised structures according to a trained structure-based energy model.
This process yields pseudo-relaxed structures and results in smaller errors in materials discovery tasks.

\clearpage
\bibliographystyle{unsrt}
\bibliography{references.bib}